\documentclass[bjps,preprint]{imsart}

\usepackage{amsmath,amssymb,epsfig}
\usepackage{subfigure}
\usepackage{graphicx}
\usepackage{epstopdf}
\usepackage{bm}
\usepackage{natbib}
\usepackage{enumerate}
\usepackage{verbatim}
\usepackage{url}

\startlocaldefs
\endlocaldefs

\newcommand{\eps}{\epsilon}

\newcommand{\be}{\begin{eqnarray}}
\newcommand{\ee}{\end{eqnarray}}
\newcommand{\ba}{\begin{eqnarray*}}
\newcommand{\ea}{\end{eqnarray*}}

\usepackage[breaklinks=true]{hyperref}
\usepackage{breakcites}
\usepackage{cite}

\begin{document}

\begin{frontmatter}

\title{Bayesian nonparametric regression using complex wavelets}
\runtitle{Bayesian nonparametric regression using complex wavelets}

\begin{aug}

\author{\fnms{Norbert} \snm{Rem\'enyi}\thanksref{a}\corref{}\ead[label=e1]{nremenyi@gatech.edu}}
\and
\author{\fnms{Brani} \snm{Vidakovic}\thanksref{b}\ead[label=e2]{brani@gatech.edu}}

\affiliation[a]{Research Group, Sabre Holdings, Southlake, Texas}
\affiliation[b]{School of Industrial and Systems Engineering, Georgia Institute of Technology}


\runauthor{Rem\'enyi and Vidakovic}

\end{aug}

\begin{abstract}
In this paper we propose a new adaptive wavelet denoising methodology using complex wavelets. The method is based on a fully Bayesian hierarchical model in the complex wavelet domain that uses a bivariate mixture prior on the wavelet coefficients. The heart of the procedure is computational, where the posterior mean is computed through Markov chain Monte Carlo (MCMC) simulations. We show that the method has good performance, as demonstrated by simulations on the well-known test functions and by comparison to a well-established complex wavelet-based denoising procedure. An application to real-life data set is also considered.

\end{abstract}



\end{frontmatter}

\section{Introduction} \label{sec_intro}
In the present paper we consider a novel Bayesian model as a solution to the classical nonparametric
regression problem
\be y_i=f(x_i)+
\varepsilon_i, \quad i=1,\dots, n,\label{problem}
\ee
where $x_i$, $i=1, \ldots,n$, are equispaced sampling points, and the errors $\varepsilon_i$ are i.i.d. normal random variables, with zero mean and variance $\sigma^2$. The interest is to estimate the function $f$ using the observations $y_i$. After applying a linear and orthogonal wavelet transform, the equation in (\ref{problem}) becomes
\be
d_{jk}=\theta_{jk}+\varepsilon_{jk} \label{model0jk},
\ee
where $d_{jk}$, $\theta_{jk}$ and $\varepsilon_{jk}$ are the wavelet coefficients (at resolution $j$ and position $k$) corresponding to $y$, $f$ and $\varepsilon$ respectively. Note that $\varepsilon_i$ and $\varepsilon_{jk}$ are equal in distribution due to orthogonality of wavelet transforms.  Due to the the whitening property of the wavelet transforms \citep{Flandrin1992} many existing methods assume independence of the coefficients, and omit the double indices $jk$ to model a generic wavelet coefficient as
\be
d=\theta+\varepsilon,\quad \eps \sim {\cal N}(0, \sigma^2). \label{model0}
\ee
When indices are needed for the clarity of exposition they will be used.

To estimate $\theta$ in model (\ref{model0}) Bayesian shrinkage rules have been proposed in the literature by many authors. By a shrinkage rule the observed wavelet coefficients $d$ are replaced with their shrunk version $\hat{\theta}=\delta(d)$. Then $f$ is estimated as the inverse wavelet transform of $\hat{\theta}$. Empirical distributions of detail wavelet coefficients for signals encountered in practical applications are (at each resolution level) centered around and peaked at zero \citep{Mallat1989}. A range of models, for which unconditional distribution of wavelet coefficients mimic these properties, have been considered in the literature. The traditional Bayesian models consider prior distribution on the wavelet coefficient
$\theta$ as
\be
\pi(\theta)=\eps\delta_0+(1-\eps)\xi(\theta) \label{mixture},
\ee
where $\delta_0$ is a point mass at zero, $\xi$ is a symmetric and unimodal distribution, and $\eps$ is a fixed parameter in [0,1], usually level dependent, that controls the extent of shrinkage for values of $d$ close to 0. This type of model was considered by \citet{Abramovich1998}, \citet{Vidakovic1998a}, \citet{Vidakovic2001} and \citet{Johnstone2005a}, among others. A recent overview of Bayesian strategies in wavelet shrinkage can be found in
\citet{RemenyiChap2012}.

Wavelet shrinkage methods using complex-valued wavelets provide additional insights to shrinkage process due to the information contained in the phase. After taking a complex wavelet transform of a real-valued signal, the model remains as in (\ref{model0jk}), however, the observed wavelet coefficients $d_{jk}$ become complex numbers at resolution $j$ and location $k$. Several papers considering Bayesian wavelet shrinkage with complex wavelets are available. \citet{Lina1995} describes the complex-valued Daubechies' wavelets in detail, \citet{Lina1997b}, \citet{Lina1997a}, and \citet{Lina1999} focus on image denoising, in which the phase of the observed wavelet coefficients is preserved, but the modulus of the coefficients is shrunk by the Bayes rule. \citet{Barber2004} develops the complex empirical Bayes (CEB) procedure, which modifies both the phase and modulus of wavelet coefficients by a bivariate shrinkage rule. The wavelet coefficients are represented as bivariate real-valued random variables and the authors formulate a bivariate model in the complex wavelet domain. The resulting estimators are in closed-form and the hyperparameters of the model are estimated by the empirical Bayes procedure.

We build on the results above, and formulate a fully Bayesian hierarchical model which accounts for the uncertainty of the prior parameters by placing hyperpriors on them. Since a closed-form solution for the Bayes estimator does not exist, the Markov Chain Monte Carlo (MCMC) methodology is applied and an approximate estimator (posterior mean) is computed from the output of simulational runs. Although the simplicity of a closed-form solution is lost, the procedure is fully Bayesian, adaptive to the underlying signal, and the estimation of the hyperparameters is automatic via the MCMC sampling algorithm. The estimation is governed by the data and hyperprior distributions on the parameters, and this adaptivity ensures good denoising performance.

The paper is organized as follows. Section \ref{sec_model} formalizes the model and presents some results related to it. Section \ref{sec_gibbs} details the MCMC sampling scheme. Section \ref{sec_sim} discusses the selection of hyperparameters and presents simulation results and comparisons to existing methods. In Section \ref{sec_data} we apply the proposed shrinkage to a real data set, and in Section \ref{sec_concl} conclusions are provided.

\section{Fully Bayesian Model} \label{sec_model}
In this section we describe a novel, fully Bayesian hierarchical model in the complex wavelet domain. After applying the complex wavelet transform to a real-valued signal, the observed wavelet coefficients $d_{jk}$ at resolution $j$ and location $k$ become complex numbers. Building on the approach taken by \citet{Barber2004}, we represent the complex-valued wavelet coefficients as bivariate real-valued random variables.

We consider the following Bayesian model
\be
d_{jk}|\theta_{jk},\sigma^2 &\sim& {\cal N}_2 (\theta_{jk}, \sigma^2\Sigma_j) \nonumber \\
\theta_{jk}|\epsilon_j,C_j  &\sim& (1-\epsilon_j)\delta_0+\epsilon_j {\cal EP}_2(\mu,C_j,\beta), 
\label{cmcmc_model1}
\ee
where ${\cal EP}_2$ stands for the bivariate exponential power distribution. The multivariate exponential power distribution is an extension of the class of normal distributions in which the heaviness of tails can be controlled. Its definition and properties can be found in \citet{Gomez1998}. The prior on the location $\theta_{jk}$ is a bivariate extension of the standard mixture prior in the Bayesian wavelet shrinkage literature, consisting of a point mass at zero and a heavy-tailed distribution. As a prior, \citet{Barber2004} considered a mixture of point mass and bivariate normal distribution. A heavy-tailed mixture prior in our proposal better captures the sparsity of wavelet coefficients common in most applications; however, a closed-form solution is lost, and we need to rely on MCMC simulations to compute estimators.

To calibrate the exponential power prior in (\ref{cmcmc_model1}) for wavelet shrinkage, we use $\mu=0$, because the detail wavelet coefficients are centered around zero by their definition. We also fix $\beta=1/2$, which gives the prior on $\theta_{jk}$ the following form:
\be
\pi(\theta_{jk}|C_j)=\frac{1}{8\pi|C_j|^{1/2}}\exp\left\{-\frac{1}{2}\left(\theta_{jk}'C_j^{-1}\theta_{jk}\right)^{1/2}\right\}. \label{cmcmc_prior}
\ee
The prior specified above is equivalent to the bivariate double exponential distribution. The univariate double exponential prior was found to have desirable properties in the real-valued wavelet denoising context \citep{Vidakovic2001,Johnstone2005a}, hence it is natural to extend it to the bivariate case.

From model (\ref{cmcmc_model1}) it is apparent that the mixture prior on $\theta_{jk}$ is set depending on the dyadic level $j$, which ensures scale-adaptivity of the method. Quantity $\sigma^2\Sigma_j$ represents the scaled covariance matrix of the noise at each decomposition level, and $C_j$ represents the levelwise scale matrix in the exponential power prior. Explicit expression for the covariance ($\Sigma_j$) induced by white noise in complex wavelet shrinkage was derived in \citet{Barber2004}. We adopt the approach described in their paper to model the covariance structure of the noise and compute $\Sigma_j$ for each dyadic level $j$ from the expressions
\be
\label{complex_cov}
\text{Cov}\{\text{Re}(\bm \varepsilon),\text{Im}(\bm \varepsilon)\} &=& -\sigma^2\text{Im}(WW^T)/2,
\nonumber \\
\text{Cov}\{\text{Re}(\bm \varepsilon),\text{Re}(\bm \varepsilon)\} &=& \sigma^2\{I_n+\text{Re}(WW^T)\}/2,  \\
\text{Cov}\{\text{Im}(\bm \varepsilon),\text{Im}(\bm \varepsilon)\} &=& \sigma^2\{I_n-\text{Re}(WW^T)\}/2.\nonumber
\ee
We assume a common i.i.d normal noise model $\bm e \sim {\cal N}_n(\bm 0,\sigma^2I_n)$, as in (\ref{problem}). After taking complex wavelet transform, the real and imaginary parts of the transformed noise $\bm \varepsilon=W\bm e$ become correlated, which is expressed by (\ref{complex_cov}) above. Note that $W$ is a complex-valued unitary matrix representing the wavelet transform, therefore $\bar{W}^TW=W\bar{W}^T=I$, where $\bar{W}$ denotes the complex conjugate of $W$.

Instead of estimating hyperparameters $\sigma^2$, $\epsilon_j$, and $C_j$ in (\ref{cmcmc_model1}) in empirical Bayes fashion, we elicit hyperprior distributions on them in a fully Bayesian manner. We specify a conjugate inverse gamma prior on the noise variance $\sigma^2$, and an inverse Wishart prior on the matrix $C_j$ describing the covariance structure of the spread prior of $\theta_{jk}$. Mixing weight $\epsilon_j$ regulates the strength of shrinkage of a wavelet coefficient to zero. We specify a ``noninformative'' uniform prior on this parameter, allowing the estimation to be governed mostly by the data.

For computational purposes, we represent the exponential power prior in (\ref{cmcmc_prior}) as a scale mixture of multivariate normal distributions, which is an essential step for efficient Monte Carlo simulation. From \citet{Gomez2008}, the bivariate exponential power distribution with $\mu=0$ and $\beta=1/2$ can be represented as
\ba
{\cal EP}_2(\mu=0,C_j,\beta=1/2)=\int_0^{\infty} {\cal N}_2(0,v C_j)\frac{1}{\Gamma(3/2)8^{3/2}}v^{1/2}e^{-v/8}dv,
\ea
which is a scale mixture of bivariate normal distributions with mixing distribution gamma. Using the specified hyperpriors and the mixture representation, the basic model in (\ref{cmcmc_model1}) extends to
\be
d_{jk}|\theta_{jk},\sigma^2     &\sim& {\cal N}_2 (\theta_{jk}, \sigma^2\Sigma_j) \nonumber \\
\sigma^2                        &\sim& {\cal IG} (a,b) \nonumber \\
\theta_{jk}|z_{jk},v_{jk},C_j   &\sim& (1-z_{jk})\delta_0+z_{jk} {\cal N}_2(0,v_{jk}C_j) \nonumber \\
z_{jk}|\epsilon_j               &\sim& {\cal B}er (\epsilon_j) \\
\epsilon_j                      &\sim& {\cal U} (0,1) \nonumber \\
v_{jk}                          &\sim& {\cal G}a (3/2,8) \nonumber \\
C_j                             &\sim& {\cal IW} (A_j,w).\nonumber  \label{cmcmc_model2}
\ee
For computational purposes, we introduce a latent variable $z_{jk}$. Variable $z_{jk}$ is a Bernoulli variable indicating whether parameter $\theta_{jk}$ comes from a point mass at zero ($z_{jk}=0$) or from a bivariate normal distribution ($z_{jk}=1$), with prior probability of $1-\eps_j$ or $\eps_j$, respectively. The uniform ${\cal U}(0,1)$ prior on $\eps_j$ is equivalent to beta ${\cal B}e(1,1)$ distribution, which is a conjugate prior for the Bernoulli distribution. Integrating out $z_{jk}$ from model (\ref{cmcmc_model2}) gives back the original mixture expression in model (\ref{cmcmc_model1}).

By representing the exponential power prior as a scale mixture of normals, the hierarchical model in (\ref{cmcmc_model2}) becomes tractable, because the full conditional distributions of all the parameters become explicit. Therefore, we can develop a fast Gibbs sampling algorithm to update all the necessary parameters $\sigma^2$, $z_{jk}$, $\eps_j$, $\theta_{jk}$, $v_{jk}$ and $C_j$. Note that in order to run the Gibbs sampling algorithm we only have to specify hyperparameters $a$, $b$, $A_j$ and $w.$  The details of Gibbs sampling scheme will be discussed in the next section.

\section{Gibbs sampling scheme} \label{sec_gibbs}
To conduct posterior inference on the wavelet coefficients $\theta_{jk}$, a standard Gibbs sampling procedure is adopted. In this section we provide details of how to develop a Gibbs sampler for the model in (\ref{cmcmc_model2}). Gibbs sampling is an iterative algorithm that simulates from a joint posterior distribution through iterative simulation over the list of full conditional distributions. For more details on Gibbs sampling, see \citet{Casella1992}, or \citet{Robert1999}. For the model in (\ref{cmcmc_model2}), the full conditionals  for parameters $\sigma^2$, $z_{jk}$, $\eps_j$, $\theta_{jk}$, $v_{jk}$ and $C_j$ can be specified exactly.
Specification of hyperparameters $a$, $b$, $A_j$ and $w$ will be discussed in Section \ref{sec_sim}.
Technical details of this section are deferred to Appendix.

\subsection{Updating $\sigma^2 $}
Using a conjugate ${\cal IG} (a,b)$ prior on $\sigma^2$ results in a full conditional which is inverse gamma. Therefore, update $\sigma^2$ as
\footnotesize
\be
{\sigma^2}^{(i)} \sim {\cal IG} \left(a+n,\left[1/b+1/2\sum_{jk}\left(d_{jk}-\theta^{(i-1)}_{jk}\right)'\Sigma_j^{-1}
\left(d_{jk}-\theta^{(i-1)}_{jk}\right)\right]^{-1} \right),
\ee
\normalsize
where $n=2^J-2^{J_0}$ denotes the sample size, and $i$ denotes the $i^{th}$ simulation run.

\subsection{Updating $z_{jk}$ and $\eps_{j}$}
In model (\ref{cmcmc_model2}) the latent variable $z_{jk}$ has Bernoulli prior with parameter $\eps_j$. Its full conditional remains Bernoulli and is updated as follows:
\footnotesize
\be
z^{(i)}_{jk}=
\begin{cases}
0, & \mbox{wp.} \quad \displaystyle
    \frac{\left(1-\eps^{(i-1)}_j\right)f\left(d_{jk}|0,{\sigma^2}^{(i)}\right)}
    {\left(1-\eps^{(i-1)}_j\right)f\left(d_{jk}|0,{\sigma^2}^{(i)}\right)+
    \eps^{(i-1)}_j m\left(d_{jk}|{\sigma^2}^{(i)},v_{jk}^{(i-1)},C_j^{(i-1)}\right)} \\ \nonumber
1, & \mbox{wp.} \quad \displaystyle
    \frac{\eps^{(i-1)}_j m\left(d_{jk}|{\sigma^2}^{(i)},v_{jk}^{(i-1)},C_j^{(i-1)}\right)}
    {\left(1-\eps^{(i-1)}_j\right)f\left(d_{jk}|0,{\sigma^2}^{(i)}\right)+
    \eps^{(i-1)}_j m\left(d_{jk}|{\sigma^2}^{(i)},v_{jk}^{(i-1)},C_j^{(i-1)}\right)}
\end{cases}, \\
\ee
\normalsize
where
\footnotesize
\ba
f(d_{jk}|0,\sigma^2) &=& \frac{1}{2\pi|\sigma^2\Sigma_j|^{1/2}}\exp\left\{-\frac{1}{2\sigma^2}d_{jk}'\Sigma_j^{-1}d_{jk}\right\}, \\
m(d_{jk}|\sigma^2,v_{jk},C_j) &=& \frac{1}{2\pi|\sigma^2\Sigma_j+v_{jk}C_j|^{1/2}}
\exp\left\{-\frac{1}{2}d_{jk}'\left(\sigma^2\Sigma_j+v_{jk}C_j\right)^{-1}d_{jk}\right\}.
\ea
\normalsize
Parameter $\eps_j$ is given a conjugate ${\cal B}e(1,1)$ prior. This results in a full conditional distributed as beta. Therefore we update $\eps_j$ as
\footnotesize
\be
\eps^{(i)}_j \sim {\cal B}e \left(1+\sum_k z^{(i)}_{jk}, 1+\sum_k \left(1-z^{(i)}_{jk}\right) \right).
\ee
\normalsize
Note that other choices from the ${\cal B}e(\alpha,\beta)$ family are possible as the prior for $\eps_j$. However, we used the noninformative choice of $\alpha=1$ and $\beta=1$ to facilitate data-driven estimation of $\eps_j$.

\subsection{Updating $\theta_{jk} $}
From the conjugate setup of model (\ref{cmcmc_model2}) and using the latent variable $z_{jk}$, it follows that the full conditional distribution of $\theta_{jk}$ is either a point mass at zero ($z_{jk}=0$), or a bivariate normal distribution ($z_{jk}=1$). Therefore we update $\theta_{jk}$ as follows:
\footnotesize
\be
\theta^{(i)}_{jk} \sim
\begin{cases}
\delta_0(\theta_{jk}), & \mbox{if} \quad z^{(i)}_{jk}=0 \\
f\left(\theta_{jk}|d_{jk},{\sigma^2}^{(i)},v_{jk}^{(i-1)},C_j^{(i-1)}\right), & \mbox{if} \quad z^{(i)}_{jk}=1
\end{cases},
\ee
\normalsize
where
\footnotesize
\ba
f(\theta_{jk}|d_{jk},\sigma^2,v_{jk},C_j) &=& \frac{1}{2\pi|\tilde{\Sigma}_{jk}|^{1/2}}\exp\left\{-\frac{1}{2}\tilde{\mu}_{jk}'\tilde{\Sigma}_{jk}^{-1}
\tilde{\mu}_{jk}\right\}, \\
\tilde{\mu}_{jk} &=& \tilde{\Sigma}_{jk} \frac{\Sigma^{-1}_j}{\sigma^2}d_{jk}, \\
\tilde{\Sigma}_{jk} &=& \left(\Sigma^{-1}_j/\sigma^2+C^{-1}_j/v_{jk}\right)^{-1}.
\ea
\normalsize

\subsection{Updating $v_{jk}$}
In model (\ref{cmcmc_model2}) for the scale mixture of normals representation
we placed a gamma prior on $v_{jk}.$
The full conditional distribution of $v_{jk}$ depends on the value of $z_{jk}$, and the updating scheme is:
\be
\footnotesize
v^{(i)}_{jk} \sim
\begin{cases}
{\cal G}a (3/2,8), & \mbox{if} \quad z^{(i)}_{jk}=0 \\
{\cal GIG} \left(1/4,{\theta^{(i)}_{jk}}'\left\{C^{(i-1)}_j\right\}^{-1}\theta^{(i)}_{jk},1/2 \right), & \mbox{if} \quad z^{(i)}_{jk}=1
\end{cases}.
\normalsize
\ee
Here ${\cal GIG}(a,b,p)$ denotes the generalized inverse Gaussian distribution \citep[p.284]{Johnson1994} with probability density function
\ba
\footnotesize
f(x|a,b,p)=\frac{(a/b)^{p/2}}{2K_p(\sqrt{ab})}x^{p-1}e^{-(ax+b/x)/2}, \quad x>0; a,b>0,
\normalsize
\ea
where $K_p$ denotes the modified Bessel function of the third kind. Simulation of ${\cal GIG}$ random variates is available through a MATLAB$^{\copyright}$ implementation based on \citet{Dagpunar1989}.

\subsection{Updating $C_j$}
Placing a conjugate inverse Wishart prior on covariance matrix $C_j$ results in a full conditional distribution which remains inverse Wishart. Therefore, $C_j$ is updated as:
\be
\footnotesize
C_j^{(i)} \sim {\cal IW}\left( A_j+\sum_{k}z^{(i)}_{jk} \frac{\theta^{(i)}_{jk}{\theta^{(i)}_{jk}}'}{v^{(i)}_{jk}}, w+\sum_{k}z^{(i)}_{jk} \right).
\normalsize
\ee
The implementation of the described Gibbs sampler requires simulation routines for standard distributions such as inverse gamma, Bernoulli, beta, normal, and also a specialized routine to simulate from the generalized inverse Gaussian. The procedure was implemented in MATLAB and available from the authors.

The Gibbs sampling procedure can be summarized as
\begin{enumerate}[(i)]
\item Choose initial values for parameters
\item Repeat steps (iii) - (viii) for $l=1,\ldots,M$
\item Update $\sigma^2$
\item Update $z_{jk}$ for $j=J_0,\ldots,\log_2(n)-1,~k=0,\ldots,2^j-1$
\item Update $\eps_j$ for $j=J_0,\ldots,\log_2(n)-1$
\item Update $\theta_{jk}$ for $j=J_0,\ldots,\log_2(n)-1,~k=0,\ldots,2^j-1$
\item Update $v_{jk}$ for $j=J_0,\ldots,\log_2(n)-1,~k=0,\ldots,2^j-1$
\item Update $C_j$ for $j=J_0,\ldots,\log_2(n)-1$.
\end{enumerate}
The denoising method based on the Gibbs sampling algorithm above will be called Complex Gibbs Sampling Wavelet Smoother (\textit{CGSWS}). In the following section we explain the specification of hyperparameters $a$, $b$, $A_j$ and $w$, and apply the {\textit{CGSWS}} algorithm to denoise simulated test functions.

\section{Simulations and Comparisons} \label{sec_sim}
In this section we discuss the performance of the proposed {\textit{CGSWS}} estimator and compare it to an established method from the literature considering complex Bayesian wavelet denoising. Within each replication of the simulations we performed 10,000 Gibbs sampling iterations, of which the first 5,000 was burn-in. We used the sample average $\hat{\theta}_{jk}=\sum_{i} \theta^{(i)}_{jk}/N$ as the usual estimator for the posterior mean. In our set-up $N=5,000$. First we discuss the selection of hyperparameters, then explain the simulation setup and results.

\subsection{Selection of Hyperparameters} \label{sec_hyper_complex_gsws}
In any Bayesian modeling task the selection of hyperparameters is critical for good performance of the model. It is also desirable to have a default way of selecting the hyperparameters which makes the shrinkage procedure automatic.

In order to apply the \textit{CGSWS} method we  need to specify hyperparameters $a$, $b$, $A_j$ and $w$ in the hyperprior distributions. This selection is governed by the data and hyperprior distributions.
The advantage of the fully Bayesian approach is that once the hyperpriors are set, the estimation of parameters $\sigma^2$, $\eps_j$, $\theta_{jk}$, $v_{jk}$ and $C_j$ is automatic via the Gibbs sampling scheme.  Another advantage is that the method is relatively robust to the choice of hyperparameters since they enter the model at higher level of hierarchy.\\

\noindent {\bf Parameters $a$ and $b$.} For simplicity we set $a=2$ and $b=1/\hat{\sigma}^2$, where
\ba
\hat{\sigma}^2=\left(\textnormal{MAD}(d^{\textnormal{re}}_{jk}/0.6745)\right)^2+
\left(\textnormal{MAD}(d^{\textnormal{im}}_{jk}/0.6745)\right)^2, \quad j=\log_2(n)-1.
\ea
This ensures that the mean of the inverse gamma prior on $\sigma^2$ is the standard robust estimator of the noise variation \citep{Donoho1994}. Here MAD stands for the median absolute deviation of the wavelet coefficients, which we calculate at the finest level of detail from both the real and imaginary parts of wavelet coefficients \citep{Barber2004}.\\

\noindent {\bf Parameters $A_j$ and $w$.} Hyperparameters $A_j$ and $w$ play an important role in the prior on the covariance matrix $C_j$. Since in the Gibbs sampler updates $\sum_{k}z^{(i)}_{jk}$ and therefore $\sum_{k}z^{(i)}_{jk}\theta^{(i)}_{jk}{\theta^{(i)}_{jk}}'/v^{(i)}_{jk}$ can possibly be zero, a noninformative Jeffreys prior on $C_j$ is not computationally feasible. Also note that the mean of the inverse Wishart prior is $A_j/(w-p-1)$, where $p$ is the dimension of $A_j$, which is equal to 2 in our case. Therefore we set
\ba
A_j=(w-2-1)\hat{C}_j,
\ea
which forces the mean of the prior to be a pre-specified estimate of $C_j$. In the case of the mixture bivariate double exponential prior, the covariance of the signal part is $\textnormal{Cov}(\theta_{jk})=\eps_j^2\,12\,C_j$, where $12\,C_j$ is the covariance of a bivariate double exponential random variable \citep{Gomez1998}. Since the model assumes independence of signal and error parts, we have that $\textnormal{Cov}(d_{jk})=\eps_j^2\, 12\, C_j+\sigma^2\Sigma_j$, where $\textnormal{Cov}(d_{jk})$ is the covariance of the observations $d_{jk}$ at $j^{th}$ dyadic level. We choose $\eps_j=1/\sqrt{12}$ as a reasonable estimate, which additionally simplifies the equation in hand. Therefore a reasonable estimator for $C_j$ is
\be
\hat{C}_j=\textnormal{Cov}(d_{j})-\hat{\sigma}^2\Sigma_j, \quad J_0\leq j \leq \log_2{n}-1, \label{cmcmc_setC}
\ee
where $\textnormal{Cov}(d_{j})$ is the sample covariance estimator using observations $d_{jk}$ at $j^{th}$ dyadic level. Note that $\Sigma_j$ is known, and $\hat{\sigma}^2$ is the usual robust estimator of the variance of wavelet coefficients introduced before. Also note that when $\hat{C}_j$ is not positive definite, we regularize it by adding a multiple of the identity matrix.

Finally, we set $w=10$. Note, that $w=4$ is the least informative choice in our case, however, we found that a slightly higher values for $w$ worked better in practice.

\subsection{Simulation Results}
In the following section we present results of a simulation study in which we compare the denoising performance of the \textit{CGSWS} method to two complex wavelet-based denoising methods introduced by \citet{Barber2004}. The first one (\textit{CMWS-Hard}) is a phase preserving estimator based on hard thresholding of a ``thresholding statistic'' $d_{jk}'\Sigma_j^{-1}d_{jk}$. The second one (\textit{CEB-Posterior mean}) is a bivariate posterior mean estimator based on an empirical Bayes procedure.

Four standard test functions ({\tt Blocks}, {\tt Bumps}, {\tt Doppler}, {\tt Heavisine}) were considered \citep{Donoho1994} in the simulations. The functions were rescaled so that the added noise produced preassigned signal-to-noise ratio (SNR), as standardly done. The test functions were simulated at $n=256$, $512$, and $1024$ equally spaced points in the interval $[0,1]$. Four commonly considered SNR's were selected, SNR=3, SNR=5, SNR=7 and SNR=10. We used the symmetric complex-valued Daubechies wavelet base with 3 vanishing moments for all the test functions. The coarsest decomposition level was $J_0=3$ which matches $J_0=\lfloor \log_2(\log(n))+1 \rfloor$ suggested by \citet{anto:01}.

Reconstruction of the theoretical signal was measured by the average mean squared error (AMSE), calculated as
\ba
\frac{1}{Mn}\sum_{k=1}^{M}\sum_{i=1}^{n}\left(\hat{f}_k(t_i)-f(t_i)\right)^2,
\ea
where $M$ is the number of simulation runs and $f(t_i),~i=1,\ldots,n$ are known values of the test functions considered. We denote by $\hat{f}_k(t_i),~i=1,\ldots,n$ the estimator from the $k$-th simulation run. Note, that in each of these simulation runs we perform 10,000 Gibbs iterations to get the estimators $\hat{\theta}_{jk}$. We set $M=100$.

The results are summarized in Table \ref{table3_gsws}, where boldface numbers indicate the
smallest AMSE result for each test scenario. The results convey that the proposed \textit{CGSWS} method outperforms both estimators for majority of the test scenarios, and in most other cases it is very close in performance to the superior method. The improvement is most pronounced at small sample sizes ($n=256$) and for the test functions \tt Bumps \rm and \tt Heavisine\rm. This result confirms the adaptiveness of the method and the advantage of using a heavy-tailed prior as prior distribution for the location of wavelet coefficients. Note, however, that the computational cost of the algorithm is higher than for the competitors. The \textit{CEB-Posterior mean} method can be a good compromise in terms of performance and computational efficiency.

\begin{table}[!htbp]
\caption[AMSE of \textit{CGSWS} method compared to estimators \textit{CMWS-Hard} and \textit{CEB-Posterior mean}.]{AMSE of \textit{CGSWS} method compared to estimators \textit{CMWS-Hard} and \textit{CEB-Posterior mean}.}
\centering
\resizebox{\linewidth}{!} {
\begin{tabular}[h]{|l|c|c||c|c|c|c|||l|c|c||c|c|c|c|}
	\hline
Signal & N & Method & SNR=3 & SNR=5 & SNR=7 & SNR=10 & Signal &  N & Method & SNR=3 & SNR=5 & SNR=7 & SNR=10 \\
	\hline \hline
Blocks     & 256  &CGSWS  &  \bf     0.4293  &  \bf     0.4533  &  \bf     0.4610  &  \bf     0.4499 &
Doppler    & 256  &CGSWS  &  \bf     0.3093  &  \bf     0.3119  &  \bf     0.3251  &  \bf     0.3619 \\
           &      &CMWS-H &          0.4929  &          0.5476  &          0.5490  &          0.5021 &
           &      &CMWS-H &          0.3332  &          0.3351  &          0.3644  &          0.4000 \\
           &      &CEB-PM &          0.4343  &          0.4675  &          0.4715  &          0.4547 &
           &      &CEB-PM &          0.3137  &          0.3158  &          0.3351  &          0.3723 \\ \hline
           & 512  &CGSWS  &  \bf     0.2954  &  \bf     0.3180  &          0.3138  &          0.3051 &
           & 512  &CGSWS  &          0.1854  &          0.2073  &          0.2052  &  \bf     0.2095 \\
           &      &CMWS-H &          0.3481  &          0.3627  &          0.3457  &          0.3166 &
           &      &CMWS-H &          0.2048  &          0.2217  &          0.2192  &          0.2289 \\
           &      &CEB-PM &          0.3028  &          0.3202  &  \bf     0.3126  &  \bf     0.2995 &
           &      &CEB-PM & \bf      0.1845  &  \bf     0.2007  &  \bf     0.2035  &          0.2132 \\ \hline
           & 1024 &CGSWS  &          0.1991  &          0.2013  &          0.1991  &          0.1924 &
           & 1024 &CGSWS  &  \bf     0.1034  &  \bf     0.1209  &          0.1310  &          0.1467 \\
           &      &CMWS-H &          0.2372  &          0.2230  &          0.2098  &          0.1944 &
           &      &CMWS-H &          0.1160  &          0.1329  &          0.1432  &          0.1601 \\
           &      &CEB-PM &  \bf     0.1980  &  \bf     0.1988  &  \bf     0.1947  &  \bf     0.1879 &
           &      &CEB-PM &          0.1087  &          0.1225  &  \bf     0.1302  &  \bf     0.1419 \\ \hline
Bumps      & 256  &CGSWS  &  \bf     0.4631  &  \bf     0.4825  &  \bf     0.4946  &  \bf     0.5181 &
Heavisine  & 256  &CGSWS  &  \bf     0.1198  &  \bf     0.1640  &  \bf     0.1900  &  \bf     0.2030 \\
           &      &CMWS-H &          0.5972  &          0.5946  &          0.5853  &          0.5809 &
           &      &CMWS-H &          0.1547  &          0.2075  &          0.2144  &          0.2198 \\
           &      &CEB-PM &          0.4855  &          0.4996  &          0.5120  &          0.5390 &
           &      &CEB-PM &          0.1338  &          0.1838  &          0.2098  &          0.2188 \\ \hline
           & 512  &CGSWS  &  \bf     0.3273  &  \bf     0.3274  &  \bf     0.3235  &  \bf     0.3203 &
           & 512  &CGSWS  &  \bf     0.0799  &  \bf     0.1050  &  \bf     0.1258  &          0.1429 \\
           &      &CMWS-H &          0.3983  &          0.3760  &          0.3538  &          0.3317 &
           &      &CMWS-H &          0.0959  &          0.1202  &          0.1357  &  \bf     0.1371 \\
           &      &CEB-PM &          0.3295  &          0.3315  &          0.3287  &          0.3228 &
           &      &CEB-PM &          0.0881  &          0.1167  &          0.1340  &          0.1427 \\ \hline
           & 1024 &CGSWS  &          0.1965  &  \bf     0.1970  &  \bf     0.2009  &  \bf     0.2090 &
           & 1024 &CGSWS  &  \bf     0.0487  &  \bf     0.0650  &  \bf     0.0747  &          0.0843 \\
           &      &CMWS-H &          0.2137  &          0.2151  &          0.2134  &          0.2223 &
           &      &CMWS-H &          0.0557  &          0.0746  &          0.0793  &  \bf     0.0791 \\
           &      &CEB-PM &  \bf     0.1919  &          0.1986  &          0.2034  &          0.2122 &
           &      &CEB-PM &          0.0564  &          0.0730  &          0.0794  &          0.0835 \\ \hline
\end{tabular}
}
\label{table3_gsws}
\end{table}

\section{Application to Inductance Plethysmography Data} \label{sec_data}
For illustration we apply the described \textit{CGSWS} method to a real-life data set from anesthesiology collected by inductance plethysmography. The recordings were made by the Department of Anaesthesia at the Bristol Royal Infirmary and represent measure of flow of air during breathing. The data set was analyzed by several authors, for example \citet{Nason1996} and \citet{Abramovich1998, Abramovich2002} where more information about the data
can be found.

\begin{figure}[!htbp]
\centering
\includegraphics[width=8cm,height=8cm]{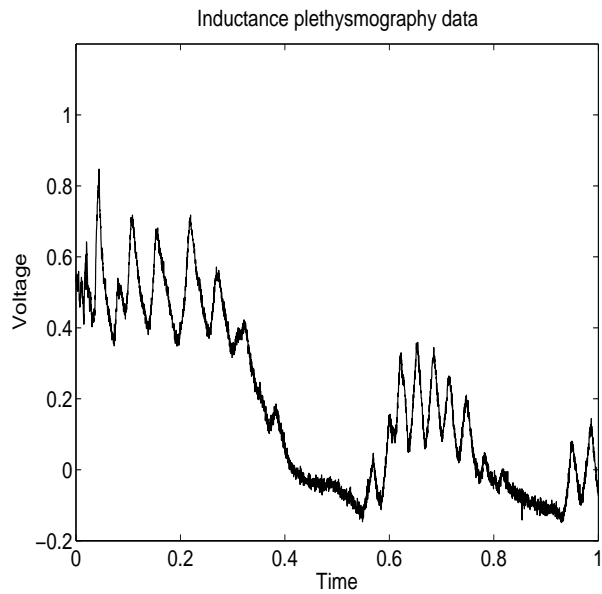}
\caption{A section of inductance plethysmography data with $n=4096$.}
\label{fig:ipd}
\end{figure}

\begin{figure}[!htbp]
\centering
\includegraphics[width=8cm,height=8cm]{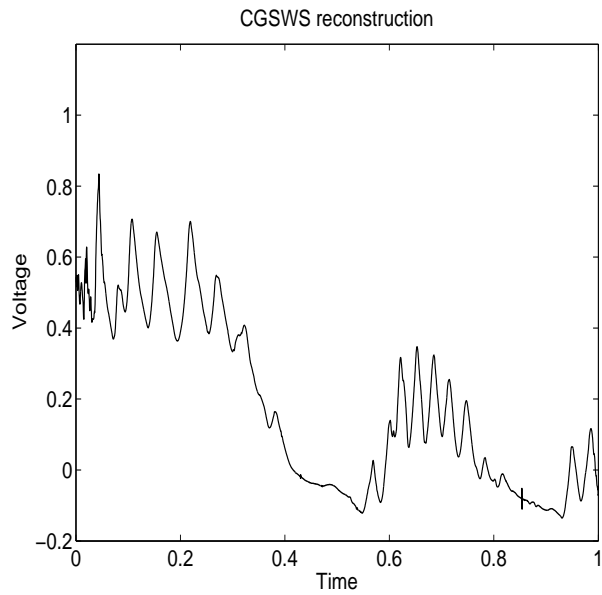}
\caption{Reconstruction of the inductance plethysmography data by \textit{CGSWS}.}
\label{fig:ipdCGSWS}
\end{figure}

Figure \ref{fig:ipd} shows a section of plethysmograph recording lasting approximately 80 s ($n=4096$ observations), while Figure \ref{fig:ipdCGSWS} shows the reconstruction of the signal with the \textit{CGSWS} method. In the reconstruction process we applied $N=10,000$ iterations of the Gibbs sampler of which the first 5,000 was burn-in. The aim of smoothing was to preserve features such as peak heights while eliminating spurious high-frequency variation. The result provided by the proposed method satisfies these requirements providing a very smooth result. \citet{Abramovich2002} report the heights of the first peak while analyzing this data set. In our case the height is 0.8342, which is quite close to the result 0.8433, obtained by \citet{Abramovich2002}, and better compared to the results obtained by other established methods analyzed in their paper.

\section{Conclusions} \label{sec_concl}
In this paper we proposed the Complex Gibbs Sampling Wavelet Smoother (\textit{CGSWS}), a complex wavelet-based method for nonparametric regression. A fully Bayesian approach was taken, in which a hierarchical model was formulated that accounts for the uncertainty of the prior parameters by placing hyperpriors on them. A mixture prior was specified on the complex wavelet coefficients with a bivariate double exponential spread distribution to account for the large wavelet coefficients. Since all the full conditional distributions were available in an explicit distributional form, an efficient Gibbs sampling estimation procedure was proposed. The \textit{CGSWS} method provided excellent denoising performance, which was demonstrated by simulations on well-known test functions and by comparison to a well-established wavelet denoising method that uses complex wavelets. The methodology was also illustrated on a real-life data set from inductance plethysmography. There the proposed method performed well in both smoothing and preserving the important features of the phenomenon.


\newpage
\section{Appendix}
In this Appendix we provide some results used for setting the Gibbs sampling algorithm in (\ref{cmcmc_model2}). To derive the full conditional distribution for a parameter of interest we start with the joint distribution of all the parameters and collect the terms which contain the desired parameter. Denote
\ba
\boldsymbol{d} &=& \{d_{jk}:j=J_0,\ldots,\log_2(n)-1,~k=0,\ldots,2^j-1\}, \\
\boldsymbol{\theta} &=& \{\theta_{jk}:j=J_0,\ldots,\log_2(n)-1,~k=0,\ldots,2^j-1\},
\ea
where $d_{jk}$ and $\theta_{jk}$ are bivariate components. Similarly,  denote
\ba
\boldsymbol{z} &=& \{z_{jk}:j=J_0,\ldots,\log_2(n)-1,~k=0,\ldots,2^j-1\}, \\
\boldsymbol{\eps} &=& \{\eps_j:j=J_0,\ldots,\log_2(n)-1\}, \\
\boldsymbol{v} &=& \{v_{jk}:j=J_0,\ldots,\log_2(n)-1,~k=0,\ldots,2^j-1\},
\ea
and
\ba
\boldsymbol{C}=\{C_j:j=J_0,\ldots,\log_2(n)-1\}
\ea
the vector containing matrices $C_j$ for resolution levels $j$. The joint distribution of the data and parameters is
\small
\ba
f(\boldsymbol{d},\boldsymbol{\theta},\boldsymbol{z},\boldsymbol{\eps},\boldsymbol{v},\sigma^2,\boldsymbol{C})&=&
\left[\prod_{j,k} \frac{1}{\sqrt{2\pi}|\sigma^2\Sigma_j|^{1/2}} \right. \cdot \\
&& \left. \exp\left\{-\frac{1}{2\sigma^2}(d_{jk}-\theta_{jk})'\Sigma_j^{-1}(d_{jk}-\theta_{jk})\right\} \right] \cdot \\
&& \frac{1}{\Gamma(a)b^a}(\sigma^2)^{-a-1}e^{-\frac{1}{\sigma^2}\frac{1}{b}}
\left[\prod_{j,k} \left\{ (1-z_{jk})\delta_0+ \right. \right. \\
&& \left. \left. z_{jk}\frac{1}{\sqrt{2\pi}|v_{jk}C_j|^{1/2}}\exp\left\{-\frac{1}{2v_{jk}}
\theta_{jk}'C_j^{-1}\theta_{jk}\right\} \right\} \right] \cdot \\
&& \left[\prod_{j,k} \eps_j^{z_{jk}}(1-\eps_j)^{(1-z_{jk})} \right]
\left[\prod_j \mbox{\bf{1}}\{0\leq\eps_j\leq1\} \right] \cdot \\
&& \left[\prod_{j,k} \frac{1}{\Gamma(3/2)8^{3/2}}v_{jk}^{3/2-1}e^{-v_{jk}/8} \right] \cdot \\
&& \left[\prod_j |C_j|^{-(w+d+1)/2}\exp\left\{-\frac{1}{2}\textnormal{tr}\left(A_j C_j^{-1}\right) \right\} \right]. \\
\ea
\normalsize

\noindent From the joint distribution, the full conditional distribution of $\sigma^2$ is
\small
\ba
\displaystyle
p(\sigma^2|\boldsymbol{\theta},\boldsymbol{d}) &\propto& \left(\frac{1}{\sigma^2}\right)^{n} \exp\left\{-\frac{1}{2\sigma^2}\sum_{j,k}(d_{jk}-\theta_{jk})'\Sigma_j^{-1}(d_{jk}-\theta_{jk})\right\}
\left(\sigma^2\right)^{-a-1}e^{-\frac{1}{\sigma^2}\frac{1}{b}} \\
&=& (\sigma^2)^{-a-n-1}\exp\left\{-\frac{1}{\sigma^2}\left(1/b+1/2\sum_{j,k}(d_{jk}-\theta_{jk})'
\Sigma_j^{-1}(d_{jk}-\theta_{jk})\right)
\right\} \\
&=& {\cal IG} \left(a+n,\left[1/b+1/2\sum_{j,k}(d_{jk}-\theta_{jk})'\Sigma_j^{-1}(d_{jk}-\theta_{jk})\right]^{-1} \right). \\
\ea
\normalsize
The conditional distribution of $z_{jk}$ remains Bernoulli with posterior success probability
\small
\ba
\displaystyle
P(z_{jk}=1|d_{jk},\sigma^2,\eps_j,v_{jk},C_j) = \frac{\eps_j m\left(d_{jk}|\sigma^2,v_{jk},C_j\right)}
    {\left(1-\eps_j\right)f\left(d_{jk}|0,\sigma^2\right)+
    \eps_j m\left(d_{jk}|\sigma^2,v_{jk},C_j\right)},
\ea
\normalsize
where
\small
\ba
f(d_{jk}|0,\sigma^2) &=& \frac{1}{2\pi|\sigma^2\Sigma_j|^{1/2}}\exp\left\{-\frac{1}{2\sigma^2}d_{jk}'\Sigma_j^{-1}d_{jk}\right\}, \\
m(d_{jk}|\sigma^2,v_{jk},C_j) &=& \frac{1}{2\pi|\sigma^2\Sigma_j+v_{jk}C_j|^{1/2}}
\exp\left\{-\frac{1}{2}d_{jk}'\left(\sigma^2\Sigma_j+v_{jk}C_j\right)^{-1}d_{jk}\right\}.
\ea
\normalsize
The marginal distribution $m(d_{jk}|\sigma^2,v_{jk},C_j)$ is a bivariate normal distribution with zero mean and covariance matrix $\sigma^2\Sigma_j+v_{jk}C_j$. This follows form conjugate multivariate normal-normal structure \citep{Lindley1972}.

The full conditional distribution of $\eps_j$ is
\ba
\displaystyle
p(\eps_j|\boldsymbol{z}) &\propto& \left[\prod_{k} \eps_j^{z_{jk}}(1-\eps_j)^{(1-z_{jk})} \right] \mbox{\bf{1}}\{0\leq\eps_j\leq1\} \\
&=& \eps_j^{\sum_k z_{jk}}(1-\eps_j)^{\sum_k(1-z_{jk})} \\
&=& {\cal B}e \left(1+\sum_k z_{jk}, 1+\sum_k \left(1-z_{jk}\right) \right). \\
\ea
The full conditional distribution of $\theta_{jk}$ is
\ba
\small
\displaystyle
p(\theta_{jk}|d_{jk},z_{jk},\sigma^2,v_{jk},C_j) &\propto& \exp\left\{-\frac{1}{2\sigma^2}
(d_{jk}-\theta_{jk})'\Sigma_j^{-1}(d_{jk}-\theta_{jk})\right\} \cdot \\
&& \left[ (1-z_{jk})\delta_0+ \right. \\
&& \left. z_{jk}\frac{1}{\sqrt{2\pi}|v_{jk}C_j|^{1/2}}\exp\left\{-\frac{1}{2v_{jk}}
\theta_{jk}'C_j^{-1}\theta_{jk}\right\} \right] \\
&=&
\begin{cases}
\delta_0(\theta_{jk}), & \mbox{if} \quad z_{jk}=0 \\
f\left(\theta_{jk}|d_{jk},\sigma^2,v_{jk},C_j\right), & \mbox{if} \quad z_{jk}=1
\end{cases},
\ea
\normalsize
where
\small
\ba
f(\theta_{jk}|d_{jk},\sigma^2,v_{jk},C_j) &=& \frac{1}{2\pi|\tilde{\Sigma}_{jk}|^{1/2}}\exp\left\{-\frac{1}{2}\tilde{\mu}_{jk}'\tilde{\Sigma}_{jk}^{-1}
\tilde{\mu}_{jk}\right\}, \\
\tilde{\mu}_{jk} &=& \tilde{\Sigma}_{jk} \frac{\Sigma^{-1}_j}{\sigma^2}d_{jk}, \\
\tilde{\Sigma}_{jk} &=& \left(\Sigma^{-1}_j/\sigma^2+C^{-1}_j/v_{jk}\right)^{-1}.
\ea
\normalsize
Derivation of $f(\theta_{jk}|d_{jk},\sigma^2,v_{jk},C_j)$ is also a standard result contained for example in \citet{Lindley1972} and was used in the wavelet shrinkage context by \citet{Barber2004}.

The full conditional distribution of $v_{jk}$ is proportional to
\small
\ba
\displaystyle
p(v_{jk}|\theta_{jk},z_{jk},C_j) &\propto& \left[ (1-z_{jk})\delta_0+ z_{jk}\frac{1}{\sqrt{2\pi}|v_{jk}C_j|^{1/2}}\exp\left\{-\frac{1}{2v_{jk}}
\theta_{jk}'C_j^{-1}\theta_{jk}\right\} \right] \cdot \\
&& v_{jk}^{3/2-1}\exp\left\{-\frac{v_{jk}}{8}\right\}.
\ea
\normalsize
For $z_{jk}=0$, this becomes
\small
\ba
\displaystyle
p(v_{jk}|\theta_{jk},z_{jk}=0,C_j) &\propto& v_{jk}^{3/2-1}\exp\left\{-\frac{v_{jk}}{8}\right\} \\
&=& {\cal G}a (3/2,8),
\ea
\normalsize
and when $z_{jk}=1$, it becomes
\small
\ba
\displaystyle
p(v_{jk}|\theta_{jk},z_{jk}=1,C_j) &\propto& \frac{1}{v_{jk}}\exp\left\{-\frac{1}{2v_{jk}}
\theta_{jk}'C_j^{-1}\theta_{jk}\right\}v_{jk}^{3/2-1}\exp\left\{-\frac{v_{jk}}{8}\right\} \\
&=& v_{jk}^{1/2-1}\exp\left\{-\frac{1}{2}\left(\frac{1}{4}v_{jk}+
\theta_{jk}'C_j^{-1}\theta_{jk}\frac{1}{v_{jk}} \right) \right\} \\
&=& {\cal GIG} \left(1/4,\theta_{jk}'C_j^{-1}\theta_{jk},1/2 \right).
\ea
\normalsize
Here ${\cal GIG}(a,b,p)$ denotes the generalized inverse Gaussian distribution \citep[p.284]{Johnson1994} with probability density function
\small
\ba
f(x|a,b,p)=\frac{(a/b)^{p/2}}{2K_p(\sqrt{ab})}x^{p-1}e^{-(ax+b/x)/2}, \quad x>0;a,b>0,
\ea
\normalsize
where $K_p$ denotes the modified Bessel function of the third kind.

Finally, the full conditional distribution of $C_j$ is given as
\small
\ba
\displaystyle
p(C_j|\boldsymbol{\theta_j},\boldsymbol{z_j},\boldsymbol{v_j}) &\propto&
\prod_{k} \left[ (1-z_{jk})\delta_0+ z_{jk}\frac{1}{\sqrt{2\pi}|v_{jk}C_j|^{1/2}}\exp\left\{-\frac{1}{2v_{jk}}
\theta_{jk}'C_j^{-1}\theta_{jk}\right\} \right] \cdot \\
&& |C_j|^{-(w+d+1)/2}\exp\left\{-\frac{1}{2}\textnormal{tr}\left(A_j C_j^{-1}\right) \right\} \\
&=& \prod_{k} \left[ (1-z_{jk})\delta_0+z_{jk}\frac{1}{\sqrt{2\pi}|v_{jk}C_j|^{1/2}} \right. \cdot \\
&& \left. \exp\left\{-\frac{1}{2} \textnormal{tr}\left(\frac{\theta_{jk}\theta_{jk}'}{v_{jk}}C_j^{-1}\right) \right\} \right] |C_j|^{-(w+d+1)/2}\exp\left\{-\frac{1}{2}\textnormal{tr}\left(A_j C_j^{-1}\right) \right\} \\
&\propto& |C_j|^{-\left(\sum_k z_{jk}+w+d+1\right)/2} \exp\left\{-\frac{1}{2}\textnormal{tr}\left( \left[A_j+\sum_{k} z_{jk}\frac{\theta_{jk}\theta_{jk}'}{v_{jk}}\right]C_j^{-1} \right) \right\} \\
&=& {\cal IW} \left( A_j+\sum_{k} z_{jk}\frac{\theta_{jk}\theta_{jk}'}{v_{jk}},
w+\sum_k z_{jk} \right),
\normalsize
\ea
where ${\cal IW}$ denotes the inverse Wishart distribution.

\end{document}